\documentclass[aps,prl,twocolumn]{revtex4-1}
\usepackage{amsmath,bm,epsfig,,subfigure}
\usepackage{color}
\usepackage[normalem]{ulem}

\def\Fbox#1{\vskip1ex\hbox to 8.5cm{\hfil\fboxsep0.3cm\fbox{%
  \parbox{8.0cm}{#1}}\hfil}\vskip1ex\noindent}  %%  {TEXT} in BOX

\newcommand{\B}[1]{{\bm{#1}}}%% Bold Roman & Greek Lower & Upper Case
\let \= \equiv \let\*\cdot \let\~\widetilde \let\-\overline

\begin{document}
\title{The Cooling Rate Dependence of the Shear Modulus of Amorphous Solids}
\author{Ashwin J., Eran Bouchbinder and Itamar Procaccia}
\affiliation{Department of Chemical Physics, The Weizmann
 Institute of Science, Rehovot 76100, Israel.}
\date{\today}
\begin{abstract}

Rapidly cooling a liquid may result in a glass transition, creating an amorphous
solid whose shear and bulk moduli are finite. Even when done with constant density, these resulting
moduli depend strongly on the rate of cooling. Understanding this phenomenon calls for analyzing separately the ``Born term" that exists also in perfectly ordered materials and the contributions of the ``excess modes" that result from glassy disorder. We show that the Born term is very insensitive to the cooling rate, and all the variation in the shear modulus is due to the excess modes. We argue that this approach provides a quantitative understanding of the cooling rate dependence of a basic linear response coefficient, i.e. the shear modulus.
\end{abstract}

\maketitle

{\bf Introduction}: The appearance of new amorphous solids like bulk metallic glasses in
contemporary technology brings about a pressing need to develop a theoretical understanding of
the effect of the protocols of preparation on the resulting properties of the obtained materials \cite{96VKB,06DLDJG}.
It was excellently expressed in \cite{95VKB} that ``Since this
so-called glass transition is essentially the falling-out of equilibrium of the system because the
typical time scale of the experiment is exceeded by the typical time scale of the relaxation
times of the system, the resulting glass can be expected to depend on the way the glass was
produced, e.g., on the cooling rate of the sample or the particulars of the cooling schedule".
Thus for example whether a bulk metallic glass will tend to fail via a shear-banding instability depends
on how it was prepared \cite{09CCM,05SF,06SF}. But given the inter-particle potential and even the density of states, can we provide a theoretical framework to predict this dependence?

In this Letter we focus on the linear elastic moduli of the produced amorphous solids, and provide
a theoretical framework to understand their dependence on the rate of cooling. Since it is known
that changing the material {\em density} has a well understood effect on the elastic moduli \cite{09LP}, we concentrate here on cooling protocols that keep the density constant \cite{05SF}. In previous work the common approach to explain the protocol dependence stressed the local motifs, be them icosahedra, tetrahedra
etc. \cite{09CCM,05SF,06SF}, but these did not provide a {\em quantitative} understanding
of the issues at stake. Rather, we will argue in this Letter that one needs to distinguish between
two mechanically significant features, one related to the volume averaged Born term (and see below for
a precise definition) that is determined
by things like density, average number of bonds per particle, strength of interactions etc., and another,
non-affine term, which is all about the degree of heterogeneity in the material. This approach
is also different from the traditional view of structure and rigidity where one tries to
explain the latter in terms of the long range correlations in the former \cite{05SF}- an impossible
goal for most glassy systems. Instead we say that what matters are average properties
related to density and compressibility and the degree of mechanical heterogeneities
in the material.

{\bf Numerical simulations}: To prepare quality data for the present discussion
we have performed 2-dimensional Molecular Dynamics simulations on a binary system which is an excellent glass former and is known to have a quasi-crystalline ground state \cite{87WSS,88LB}. Each atom in the system is labeled as either ``small''(S) or ``large''(L) and all the particles interact via Lennard
Jones (LJ) potential. All distances $|\B r_i-\B r_j|$ are normalized by $\lambda_{SL}$, the distance at which the LJ potential between the two species becomes zero and
the energy is normalized by $\epsilon_{SL}$ which is the interaction energy between two species.
Temperature was measured in units of $\epsilon_{SL}/k_B$ where $k_B$ is Boltzmann's constant. For detailed information on the model potential and
its properties, we refer the reader to Ref \cite{87WSS}. The number of particles in our simulations is varying between 400 to 10000 at a number density $n = 0.985$
with a particle ratio $N_L/N_s = (1+\sqrt{5})/4$. The mode coupling temperature $T_{MCT}$ for this system is known to reside close to 0.325. All particles have identical mass $m_0=1$ and time is normalized to $ t_0 = \sqrt{\epsilon_{SL}{\lambda_{SL}}^2/m_0}$.
For the sake of computational efficiency, the interaction potential is smoothly truncated to zero along with its first two derivatives at a cut-off
distance $r_c = 2.5$. To prepare the glasses, we first start from a well equilibrated liquid at a high temperature of $T=1.2$ which is supercooled to $T=0.35$ at a quenching rate of $3.4 \times 10^{-3}$. Secondly, we then equilibrate these supercooled liquids for times greater than $20\tau_{\rm rel}$, where $\tau_{\rm rel}$ is the  time taken for the self intermediate scattering function to become $1\%$ of its initial
value. Lastly, following this equilibration, we quench these supercooled liquids deep into the glassy regime at a temperature of $T=0.01$ at various quench rates, from instantaneous to infinitely slow.
The intermediate quench rates were $3.2\times 10^{-2}\cdots 3.2\times 10^{-7}$ in jumps of one order of magnitude. The infinitely fast quench was achieved by a conjugate gradient energy minimization of the equilibrated liquid at $T=0.35$. The infinitely slow rate was replaced by taking the quasi-crystalline ground state as the reference state. One should appreciate that the slowest quench rate required 0.21 billion
MD steps which translated to 7 days of CPU time for the largest system.
%%%%%%%%%%%%%%%%%%%%%%%%%%%%%%%%%%%%%%%%%%%%%%%%%%%%%%%%%%%%%%%%%%%%%%%%%%%%%%%%%%%%%%%%5
\begin{figure}
\includegraphics[scale = 0.25]{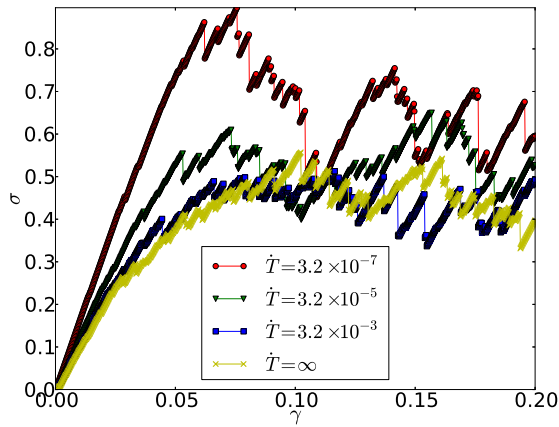}
\caption{(Color Online). Typical stress vs. strain curve obtained for AQS
straining of one realization of a system of $N$=10000 particles but for different rates of quench, as presented
in the inset. Note the significant change in the shear modulus (the slope at $\gamma=0$)
and of the yield peak stress where the system yields to plastic flow.}
\label{sigvsgam}
\end{figure}
%%%%%%%%%%%%%%%%%%%%%%%%%%%%%%%%%%%

\begin{figure}
\includegraphics[scale = 0.25]{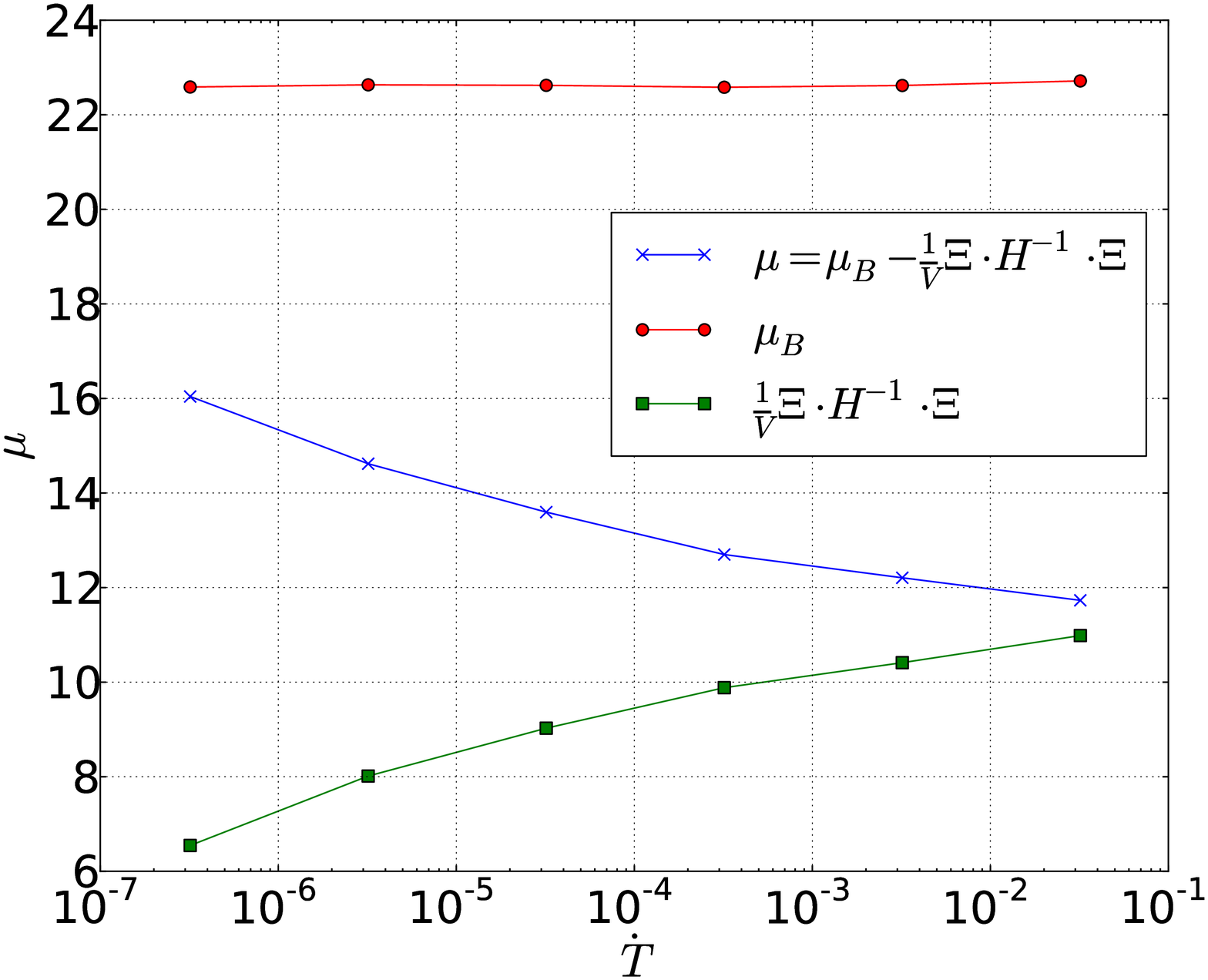}
\caption{(Color Online). Shear modulus $\mu$ at $\gamma=0$ as a function of quench rate. (N = 4900) In blue symbols $\times$
we show the shear modulus $\mu$ itself. The Born contribution is given by the red round dots.
The nonaffine contribution is given in terms of the green squares. Of course the shear modulus itself
is the difference between the two other terms.}
\label{mu}
\end{figure}

Once we have the quenched solids we can strain them using an athermal quasi-static (AQS) protocol to examine
their stress vs. strain curves. In each step of this procedure the particle positions in the system are first
changed by the affine transformation
\begin{equation}
x_i\to x_i+\delta \gamma y_i; \quad y_i\to y_i \ .
\end{equation}
This transformation results in the system not being in mechanical equilibrium, and we therefore allow
the second step, a
non affine transformation $\B r_i\to \B r_i+\B u_i$ which annuls the forces between the particles,
returning the system to mechanical equilibrium. One should understand that the non-affine transformation is
a direct result of the amorphous nature of the material: in a regular lattice without defects the affine step would leave the particles in mechanical equilibrium. The resulting data for some representative quench rates are shown in Fig. \ref{sigvsgam}. We observe that both the shear modulus and the yield peak stress (where the system yields to plastic flow) decrease significantly when the
quench rate is increased. In Fig. \ref{mu} we present the shear modulus as a function of quench rate (see
blue "$\times$" symbols"). This is the phenomenon that we want to clarify in a quantitative way.

%%%%%%%%%%%%%%%%%%%%%%%%%%%%%%%%%%%%%%%%%%%%%%%%%%%%%%%%%%%%%%%%%%%%%%%%%%%%%%%%%%%%%%%%%%%%%%

{\bf Theory}: By definition the shear modulus is the second derivative of the energy of the system
with respect to the strain $\gamma$, i.e.
\begin{equation}
\mu = \frac{1}{V}\frac{d^2 U(\B r_1,\cdots,\B r_N;\gamma)}{d \gamma^2} \ .
\end{equation}
In our process the full derivative with respect to $\gamma$ needs to be elaborated, Physically it is computed keeping the net forces
zero on all the particles, since we move from one mechanical equilibrium state to another. Thus the derivative contains two contribution, one the
partial derivative with respect to $\gamma$ and the other, via the chain rule, the contribution due to the non-affine part of the transformation \cite{72Wal,89Lut,02WTBL,04ML,04LM,10KLP}:
\begin{equation}
\frac{d}{d\gamma} = \frac{\partial}{\partial \gamma} + \frac{\partial}{\partial \B u_i} \cdot \frac{\partial \B u_i}{\partial \gamma} \equiv  \frac{\partial}{\partial \gamma} + \frac{\partial}{\partial \B r_i} \cdot \frac{\partial \B u_i}{\partial \gamma}\ ,
\end{equation}
where the second equality follows from the form of the non-affine transformation where $d\B r_i=d\B u_i$.
Applying this rule to the definition of $\mu$ we end up with the exact expression \cite{04ML,04LM,10KLP}
\begin{equation}
\mu=\frac{1}{V}\frac{\partial^2 U(\B r_1,\cdots,\B r_N;\gamma)}{\partial \gamma^2}-\frac{1}{V}
\B \Xi\cdot {\B H}^{-1} \cdot \B \Xi \ , \label{defmu}
\end{equation}
where the first term is the well known Born contribution which we denote below as $\mu_B$. The second term exists only due to the non-affine
displacement $\B u_i$ and it includes the Hessian matrix $\B H$ and the non affine ``force" $\B \Xi$ \cite{10KLP}:
\begin{equation}
H_{ij} \equiv \frac{\partial^2 U(\B r_1,\cdots,\B r_N;\gamma)}{\partial \B r_i \partial \B r_j}\ , \quad
\Xi_i \equiv \frac{\partial^2 U(\B r_1,\cdots,\B r_N)}{\partial \B r_i \partial \gamma} \ .
\end{equation}
Needless to say, before we compute the non-affine contribution in Eq.~(\ref{defmu}) we need to remove the two Goldstone modes
with $\lambda=0$ which are the result of translation symmetry.

It is very important to stress at this point that the separation between the Born term and the
non-affine term is not an arbitrary one. The Born term is very insensitive to the quench rate in our
example, and this is usually the case: it is only sensitive to average properties like density,
average number of neighbors and interactions \cite{65Zwa}. In Fig. \ref{mu} we show the result of calculating
the Born term for all our samples as a function of the quench rate, (see red dots in Fig. \ref{mu}), and there is only minor dependence. This is not the case for the non-affine term, whose direct calculation is also shown in the same figure in green squares. We see that this term changes significantly, taking upon itself the full blame of the change in the shear modulus as a function of quench rate. The sum of the two terms agrees to very high accuracy with the direct measurement of the shear modulus from the slopes of the curves
in Fig. \ref{sigvsgam} at $\gamma=0$.

\begin{figure}
\includegraphics[scale = 0.24]{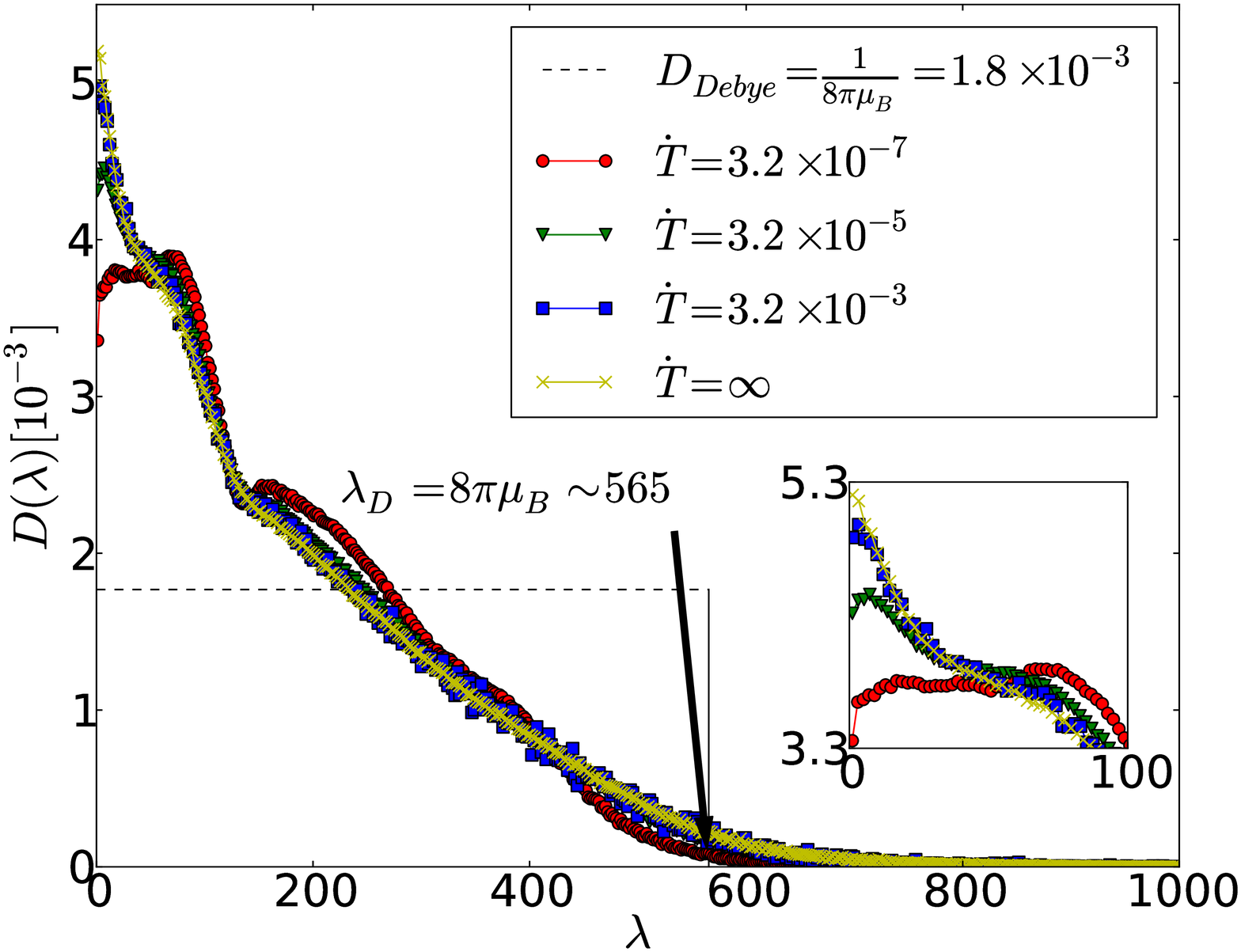}
\includegraphics[scale = 0.24]{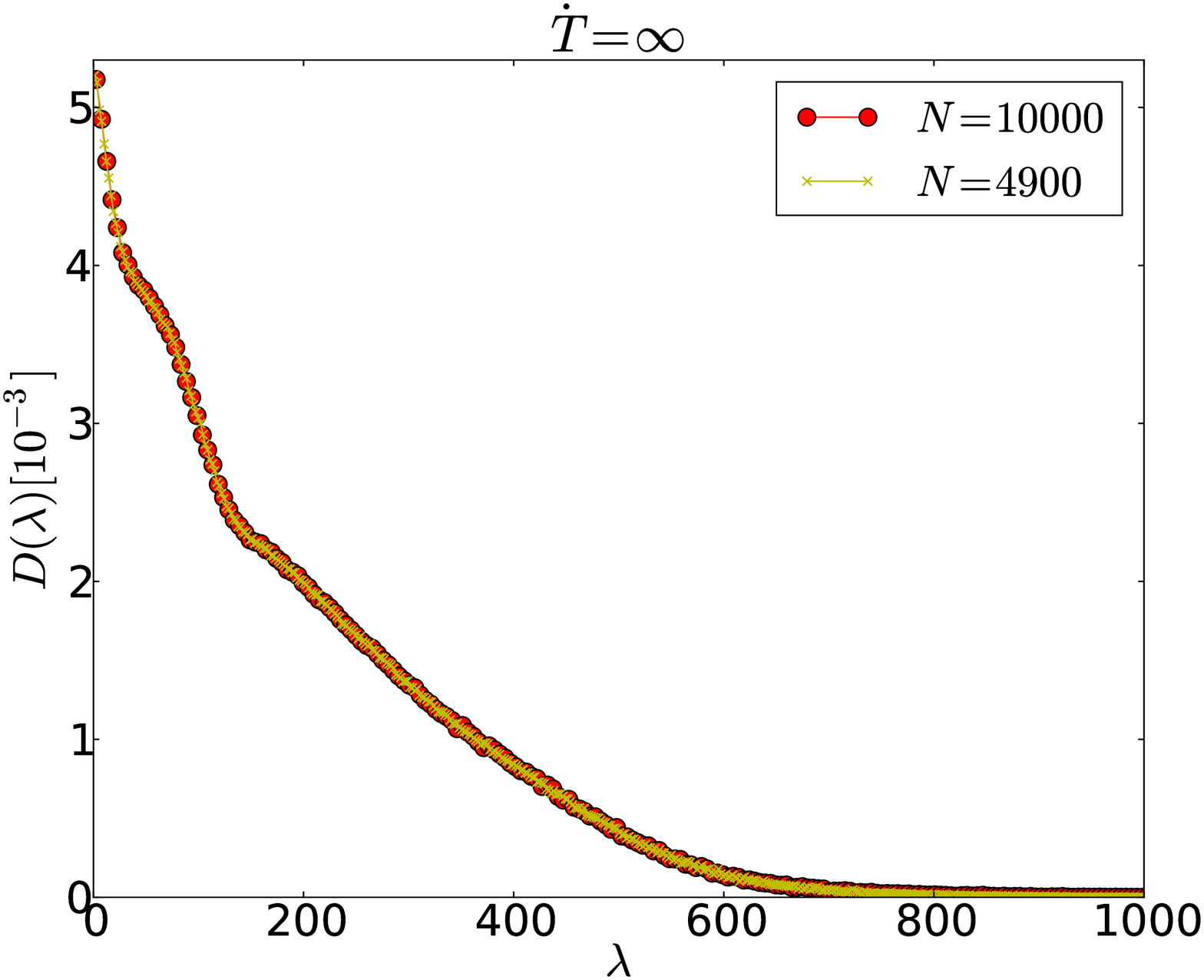}
\caption{(Color Online). Upper panel:  Density of eigenvalues of the Hessian matrix as a function of quench rate, normalized to unity.
Data was collected in bins of size $\delta\lambda=2.1$. The Debye cutoff frequency $\lambda_D$ is shown by the arrow. In the inset we show the density of states for $\lambda<100$, a region that the non-affine term in the shear modulus is very sensitive to. Lower panel: Density of eigenvalues of the Hessian matrix for the instantaneous quench at two different system sizes $N=4900$ and $N=10000$. Note that this has the highest $D(\lambda)$ at $\lambda \to 0$ and that the density of states remains stable when the system size is changed.}
\label{dos}
\end{figure}

{\bf The density of States}:
as said, the Born term is almost independent of the quench rate, and its value is very close to that
of the reference state which is the quasi-crystalline ground state. To understand the non-affine term
we need to focus now on the density of states $D(\lambda)$ where $\lambda_i$ are the eigenvalues of the Hessian matrix. For a purely elastic piece of matter lacking of any disorder we know that the
density of states is determined by the Debye theory, and in terms of the eigenvalues of the Hessian
matrix we expect a constant density
\begin{equation}
D(\lambda) = \frac{1}{8\pi \mu_B}\ , \quad \text{for a purely elastic medium}.
\end{equation}
For all our finite quench rates we have disorder in the resulting solid, and accordingly we expect
to see excess modes at small values of $\lambda$ \cite{05LBTWB,12DML}. These modes are sometime referred to as the Boson peak \cite{09IPRS}.
Their density of states is shown in Fig. \ref{dos} as a function of the quench rate. We see very clearly that the density of excess modes increases near $\lambda=0$ as the quench rate is increased. For
comparison we also show in the upper panel of Fig. \ref{dos} the constant density of states of a reference elastic medium. Since the non-affine term in Eq. \ref{defmu} has the {\em inverse} of the Hessian, any increase in the density of states near $\lambda\to 0$ should have
a strong effect on the shear modulus as is shown by the direct calculation.
%%%%%%%%%%%%%%%%%%%%%%%%%%%%%%%%%%%%%%%%%%%%%%
It is important to realize that the density of excess modes does not depend on the system size, and see
for example the lower panel of Fig. \ref{dos} in which the density of states for the same cooling rate but for two
different system sizes are superimposed. Thus one expects the same excess modes also in the thermodynamic
limit, and below we will see why the shear modulus computed in our small systems remains the same
when $N \to \infty$.

\begin{figure}
\hskip -1.5 cm
\includegraphics[scale = 0.27]{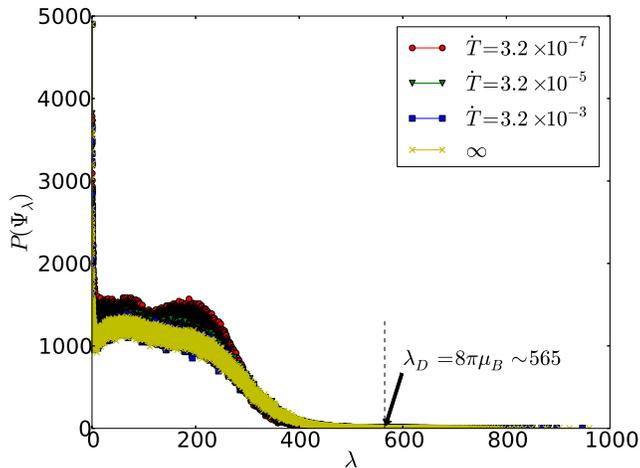}
\caption{(Color Online). The participation ratio of the mode $\Psi_\lambda$ with eigenvalue $\lambda$
at four different cooling rates as shown in the inset. The system size is $N=4900$, leading to 9800 different
modes. Note that contrary to Fig. \ref{dos} here the $\lambda$-axis was not binned. }
\label{pr}
\end{figure}

A relevant characteristic of these modes is their participation ratio, which provides a feeling as
to how extended or localized the modes are. Denoting the eigenvector associated with an eigenvalue
$\lambda_i$ as $\B \Psi_i$, we use the following definition of the participation ratio:
\begin{equation}
P(\B \Psi_i) \equiv \Big[\sum_{j=1}^N \B |\B \Psi_i^{(j)}|^2\Big]^{-1}
\end{equation}
where $\B \Psi_i^{(j)}$ is the $i$th eigenvector projected on the $j$th particle. For fully extended modes
this number is of $O(N)$ whereas for localized modes it can be much smaller. In Fig. \ref{pr} we
show the participation ratio of the modes obtained at a four different cooling rates.
%%%%%%%%%%%%%%%%%%%%%%%%%%%%%
\begin{figure}
\hskip -1.5 cm
\includegraphics[scale = 0.27]{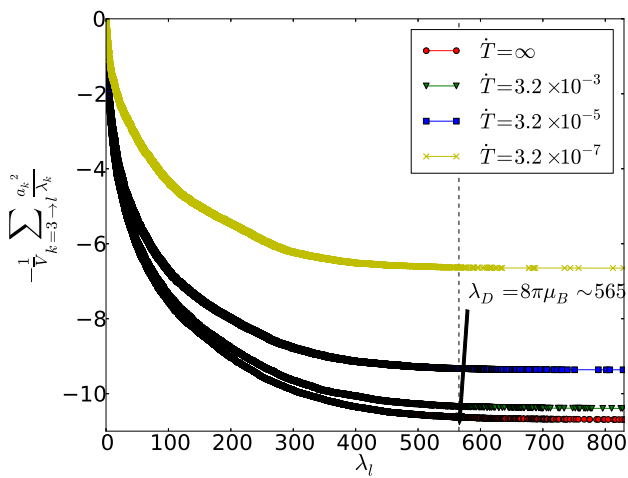}
\caption{(Color Online). The convergence of the non-affine term in the shear modulus as a function of the
upper cutoff in the sum over the modes. The calculation here is for a single realization.}
\label{converge}
\end{figure}
%%%%%%%%%%%%%%%%%%%%%%%%%%%%%%%%
We see that the very first modes, including the Goldstone modes, have a participation ratio of the order
of $N$. There is a little dip before a quasi-plateau. The modes associated with this dip are sometime
referred to as the "Quasi-Localized Modes" (QLM). Lastly there are high eigenvalue modes which are
very localized - these are the modes associated with Anderson localization.

{\bf Which modes contribute?}: the major question of interest for us at this point is which modes should be
considered to quantitatively account for the non-affine
part of the shear modulus. It had been conjectured that maybe the QLM might contribute a dominant
contribution \cite{12DML,91SL,10XVLN}. Next we provide a quantitative discussion of this issue. While the region near $\lambda \to 0$ is important, it is not sufficient to account for the full non-affine contribution. In fact all the modes up to the Debye cutoff are necessary to saturate the value of the non-affine term in the shear modulus.

In Fig. \ref{converge} we present the computed non-affine term in the shear modulus where we
use all the modes up to $\lambda_\ell$, excluding the Goldstone modes. In other words,
we compute $V^{-1}\sum_{k=3}^\ell a_k^2/\lambda_k$ where $a_k\equiv \B \Psi_k\cdot \B \Xi$. The dependence on
$\lambda_\ell$ is approximately exponential with a typical scale of 130. Thus summing up to this range of $\lambda$ yields about 63\% of the wanted quantity. To achieve an accuracy of 99\% one needs to sum up to the
Debye cutoff. All the excess modes are necessary to get the right answer. On the other
hand we see that since the density of states does not change with the system size, the calculation
of the shear modulus via this method will provide the correct shear modulus that pertains to the
thermodynamic limit even though our systems are very small.

We note that connections between the Boson peak and softening of the amorphous material were mentioned before
\cite{12DML,02TWLB,05LBTWB,91SL,08ST}. These connections did not produce however a quantitative statement of the type presented here. The obvious advantage of the present approach is its generality. The separation between the Born term
and the non-affine term is natural and robust, applying equally well to any example of amorphous solid.
The conclusion of this study is that when one sees a strong dependence of shear modulus one should seek the explanation in the non-affine rather than in the Born term. To remove any doubt that
this important conclusion is not model-dependent we repeated the present study for the very different Kob-Andersen
model \cite{93KA} and the purely repulsive model \cite{11HKLP}and found again that the Born term is highly insensitive to the cooling rate. The non-affine
term is determined by the low lying eigenvalues of the Hessian, but ``low-lying" does
not necessarily means the $\lambda\to 0$ range or even the QLM's. An accurately converged calculation
of the shear modulus require all the excess modes up to the Debye cutoff. It is possible however that
higher order elastic coefficients may require a smaller range of eigenfunctions since they are more
singular in terms of the inverse of the Hessian.

Acknowledgements: Discussions with Peter Harrowell are gratefully acknowledged. This work was supported by the Israel Science Foundation, the German-Israeli Foundation,
by the ERC under the STANPAS ``ideas" grant, the Minerva
Foundation, Munich, Germany, the Harold Perlman
Family Foundation, and the William Z. and Eda Bess
Novick Young Scientist Fund (E.B.).


\begin{thebibliography}{99}

\bibitem{96VKB}
K. Vollmayer, W. Kob and K. Binder, Phys. Rev. B {\bf 54}, 15808 (1996).

\bibitem{06DLDJG}
G. Duan, M. L. Lind, M.D. Demetriou, W. L. Johnson, W. A. Goddard,  T. Ça\'gn3, and K. Samwer, Appl. Phys. Lett. {\bf 89}, 151901 (2006).

\bibitem{95VKB}
K. Vollmayer, W. Kob and K. Binder, Europhys. Lett., {\bf 32}, 715 (1995).

\bibitem{09CCM}
A.J. Cao , Y.Q. Cheng and E. Ma, Acta Materialia {\bf 5}, 5146 (2009).

\bibitem{05SF}
Y. Shi and M. Falk, Phys. Rev. Lett. {\bf 95}, 095502 (2005).

\bibitem{06SF}
Y. Shi and M. Falk, Phys. Rev. B. {\bf 73}, 214201 (2006).


\bibitem{09LP}
E. Lerner and I. Procaccia, Phys. Rev. E, {\bf 80}, 026128 (2009).

\bibitem{87WSS}
M. Widom, K. J. Strandburg, and R. H. Swendsen, Phys. Rev. Lett. {\bf 58}, 706 (1987).

 \bibitem{88LB}
 F. Lançon and L. Billard, J. Phys. France {\bf 49}, 249 (1988).

 \bibitem{72Wal}
 D. C. Wallace, Thermodynamics of Crystals, (Wiley, New York, 1972).

 \bibitem{89Lut}
  J. F. Lutsko, J. Appl. Phys. {\bf 65}, 2991 (1989).

  \bibitem{02WTBL}
  J. P. Wittmer, A. Tanguy, J. L. Barrat, and L. Lewis, Europhys. Lett. {\bf 57}, 423 (2002).

  \bibitem{04ML}
  C. Maloney and A. Lemaitre, Phys. Rev. Lett. {\bf 93}, 195501 (2004).

  \bibitem{04LM}
  A. Lemaitre and Craig Maloney: arXiv:cond-mat/0410592v3.

 \bibitem{10KLP}
 S. Karmakar, E. Lerner, and I. Procaccia,  Phys. Rev. E 82, 026105 (2010). For a fuller detailed exposition see arXiv:1004.2198.

\bibitem{65Zwa}
R. Zwanzig and R.D. Mountain, J. Chem. Phys. {\bf 43}, 4464 (1965).

\bibitem{02TWLB}
A. Tanguy, J.P. Wittmer, F. Leonforte and J.-L, Barrat, Phys. Rev. B {\bf 66}, 174206 (2002).

\bibitem{05LBTWB}
F. Leonforte, R. Boissi\`ere, A. Tanguy, J.P. Wittmer and J.-L. Barrat, Phys. Rev. B {\bf 72}, 224206 (2005).

\bibitem{12DML}
P.M. Derlet, R. Maass and J. F. L\"offler, Eur. Phys. J. B {\bf 85}, 148 (2012).

\bibitem{09IPRS}
V. Ilyin, I. Procaccia, I. Regev, and Y. Shokef, Phys. Rev. B {\bf 80}, 174201 (2009).

\bibitem{91SL}
H.R. Schober and B.B. Laird, Phys. Rev. B {\bf 44}, 6746 (1991).


\bibitem{10XVLN}
N. Xu, V. Vitelli, A.J. Liu and S. Nagel, Euro. Phys. Lett. {\bf 90}, 56001 (2010).

\bibitem{93KA}
W. Kob and H. C. Andersen, Phys. Rev. E 48, 4364 (1993).

\bibitem{08ST}
H. Shintani and H. Tanaka, Nature Materials {\bf 7}, 870 (2008).

\bibitem{11HKLP}
H.G.E. Hentschel, S. Karmakar, E. Lerner and I. Procaccia,  Phys. Rev. E {\bf 83}, 061101 (2011).

\end{thebibliography}
\end{document}